\documentstyle[12pt,a4wide]{article}
\input epsf
\renewcommand{\theequation}{\arabic{section}.\arabic{equation}}
\newcommand{\news}{\setcounter{equation}{0}}
\newcommand{\be}{\begin{equation}}
\newcommand{\ee}{\end{equation}}
\newcommand{\bea}{\begin{eqnarray}}
\newcommand{\eea}{\end{eqnarray}}
\newcommand{\bean}{\begin{eqnarray*}}
\newcommand{\rhobf}{\mbox{\boldmath $\rho$}}
\newcommand{\eean}{\end{eqnarray*}}
\font\upright=cmu10 scaled\magstep1
\font\sans=cmss12
\newcommand{\ssf}{\sans}
\newcommand{\stroke}{\vrule height8pt width0.4pt depth-0.1pt}
\newcommand{\Z}{\hbox{\upright\rlap{\ssf Z}\kern 2.7pt {\ssf Z}}}

\newcommand{\C}{{\rlap{\rlap{C}\kern 3.8pt\stroke}\phantom{C}}}
\newcommand{\R}{\hbox{\upright\rlap{I}\kern 1.7pt R}}
\newcommand{\CP}{\C{\upright\rlap{I}\kern 1.5pt P}}
\newcommand{\PP}{\hbox{\upright\rlap{I}\kern 1.5pt P}}
\newcommand{\half}{\frac{1}{2}}
\newcommand{\mt}{\rlap{\ssf T}\kern 3.0pt{\ssf T}}
\newcommand{\spc}{spectral curve }

\newcommand{\identity}{{\upright\rlap{1}\kern 2.0pt 1}}

\begin{document}
\pagestyle{plain}

\title{\vskip -70pt
\begin{flushright}
{\normalsize DAMTP 96-32,\  UKC/IMS/96-24} \\
{\normalsize To appear in Nonlinearity} \\
\end{flushright}
\vskip 20pt
{\bf \Large \bf Inversion symmetric 3-monopoles
and the Atiyah-Hitchin manifold}
 \vskip 10pt
}
\author{Conor J. Houghton$^{\ \dagger}$
and 
Paul M. Sutcliffe$^{\ \ddagger}$
\thanks{This work was supported in part by the 
Nuffield  Foundation}\\[10pt]
{\normalsize
$\dagger$ {\sl Department of Applied Mathematics 
and Theoretical Physics}}\\
{\normalsize {\sl University of Cambridge, Silver St.,
 Cambridge CB3 9EW, England.}}\\
{\normalsize {\sl Email C.J.Houghton@damtp.cam.ac.uk}}\\[10pt]
{\normalsize 
$\ddagger$ {\sl
Institute of Mathematics,
University of Kent at Canterbury,}}\\
{\normalsize {\sl Canterbury CT2 7NZ, England.}}\\
{\normalsize {\sl Email P.M.Sutcliffe@ukc.ac.uk}}\\[10pt]
}

\date{March 1996}
\maketitle
\begin{abstract}
We consider 3-monopoles symmetric under inversion symmetry.
We show that the moduli
space of these monopoles is an Atiyah-Hitchin submanifold of the
3-monopole moduli space. 
This allows what is known about 2-monopole dynamics to be
translated into results about the dynamics of 3-monopoles.
Using a numerical ADHMN construction we compute the
monopole energy density at various points on two 
interesting geodesics. The first is a geodesic over the
two-dimensional rounded cone submanifold corresponding
to right angle scattering and the second is a closed
geodesic for three orbiting monopoles.

\end{abstract}
\newpage
\section{Introduction}
\news

The moduli space of 2-monopoles, $M_2$,  is well
understood. Its non-trivial structure is contained in the totally geodesic
submanifold of strongly centred 2-monopoles, $M_2^0$.  $M_2^0$ is the space of gauge inequivalent 2-monopoles with
fixed centre of mass and fixed overall phase. It is a hyper-K\"ahler
 4-manifold and has an $SO(3)$ action which
permutes the almost complex structures $I$, $J$ and $K$. These
properties allowed Atiyah and Hitchin to calculate the metric on
$M_2^0$ \cite{AH}
and ${M}_2^0$ is now known as the Atiyah-Hitchin manifold. 

Similarly, the non-trivial structure of the moduli space of
3-monopoles in contained in the totally geodesic submanifold of
strongly centred 3-monopoles $M_3^0$.  $M_3^0$ is also a
hyper-K\"ahler
 manifold with an $SO(3)$
action. However,
$M^3_0$ is an 8-dimensional manifold and
 these properties are not
sufficient to calculate its metric, which still remains unknown.
Although we do not know how to compute this metric, we present a more
modest result 
in this paper by proving that
there is a 4-dimensional submanifold whose metric is the
Atiyah-Hitchin one. This submanifold is the submanifold of strongly
centred 3-monopoles
which are symmetric under the inversion
\be I:(x_1,x_2,x_3)\mapsto(-x_1,-x_2,-x_3)
\label{definv}.\ee
This gives a group action on the moduli space commuting with the
$SO(3)$ action. We consider the 4-dimensional fixed point set of $I$ in
the moduli space.

\section{Monopoles}
\news
Here we are concerned with Bogomolny-Prasad-Sommerfield (BPS)
monopoles.  
A BPS monopole is a pair $(A,\Phi)$ where $A$
is a 1-form on $\R^3$ with values in $su(2)$ and $\Phi$, the Higgs
field, is an $su(2)$ valued function. They satisfy the Bogomolny equation
\be \nabla_A\Phi=\star F_A,\label{beqn}\ee
the finite energy condition 
\be \int |F_A|^2<\infty\ee
and the boundary condition
\be |\Phi|\stackrel{r\rightarrow\infty}{\longrightarrow}1.\label{bcond}\ee
The $\star$ is the Hodge star on $\R^3$ and
$\nabla_A\Phi=d\Phi+[A,\Phi]$ is the covariant derivative of
$\Phi$. An element $g$ of the gauge group $SU(2)$ acts by
\be(A,\Phi)\mapsto(gAg^{-1}-dgg^{-1},g\Phi g^{-1})\ee
and monopoles are considered equivalent if they are related by a gauge
transformation. The moduli space is the space of
gauge inequivalent monopoles. 

It can be demonstrated \cite{JT} that the
Higgs field of a monopole automatically satisfies a stronger boundary condition than
(\ref{bcond}). In fact,
\be |\Phi |=1-\frac{n}{2r}+O(r^{-2})\ee
where $n$ is an integer 
 topological charge so that the moduli space is divided up
into topological sectors. We shall call a monopole configuration with
charge $n$ an $n$-monopole. The 1-monopole is spherically symmetric and
its moduli space is $\R^3\times$S$^1$, corresponding to a position and
internal phase. The phase can be gauge transformed to unity, but it is
convenient to consider it as a degree of freedom for the
1-monopole. The position of a 1-monopole is well defined and can be
taken to be the position of the unique zero of the Higgs field.

To precisely define the moduli space of
$n$-monopoles we follow \cite{HMM} and
first define a framed monopole \cite{AH}. We say a monopole $(A,\Phi)$ is
framed if
\be
\lim_{x_3\rightarrow\infty}\Phi(0,0,x_3)=\left(\begin{array}{cc}i&0\\0&-i\end{array}\right).\ee
Every monopole can be gauge transformed into a framed one. We
define a framed gauge transformation as one for which 
\be\lim_{x_3\rightarrow\infty}g(0,0,x_3)=1.\ee
The $n$-monopole moduli space $M_n$ is now the quotient of the space
of framed monopoles by the space of framed gauge transformations.

It is more difficult to define the moduli space $M_n^0$ of strongly
centred $n$-monopoles. To do this precisely we must first discuss
rational maps. We will discuss rational maps in the next
Section. Roughly the space of strongly centred $n$-monopoles is the
submanifold of $M_n$ of monopoles with the centre of mass fixed at
the origin and the overall internal phase factor fixed at one. This
manifold is $(4n-4)$-dimensional. Over most of the manifold the $4n-4$
coordinates can be understood as corresponding to the relative
positions and relative phases of $n$ well-separated monopoles.

There is a natural metric on $M_n^0$. It
is derived from the $L^2$-norm on the fields $(A,\Phi)$. This space is
known to be hyper-K\"ahler for all $n$. This means that there is a trio of
 complex structures on $M_n^0$ satisfying the Hamilton relation
$IJ=K$. The full moduli space $M_n$ is also hyper-K\"ahler and, in fact, there
is an isometric splitting
\be \widetilde{M}_n=\R^3\times\mbox{S}^1\times M_n^0\ee
where $\widetilde{M}_n$ is an $n$-fold covering of $M_n$.

The Bogomolny equation is
time independent and its solutions are
static solutions to a Yang-Mills Higgs field theory in 
(3+1)-dimensions. The dynamics
of monopoles in this theory is of interest and the low energy dynamics 
can be approximated by geodesic flow in
the moduli space of BPS monopoles \cite{M2,St}.

\section{Spectral curves, Nahm data and rational maps}
\news

It is difficult to study monopoles directly by solving the Bogomolny
equation (\ref{beqn}) for the fields $A$ and $\Phi$. However, there are
powerful mathematical correspondences
which allow us to study monopoles indirectly. To be precise, there is a
correspondence between $n$-monopoles and the following \cite{Hb,N,D}\\

\noindent {\bf A}. Spectral curves.

 A spectral curve is an algebraic
curve $S \subset T\PP_1$ which has the form
\be \eta^n+\eta^{n-1} a_1(\zeta)+\ldots+\eta^r a_{n-r}(\zeta)+
\ldots+\eta a_{n-1}(\zeta)+a_n(\zeta)=0\label{algcurve}
\label{gensc}
\ee
where, for $1\leq r\leq n$, $a_r(\zeta)$ is a polynomial in $\zeta$ of
degree not greater than $2r$. Here $\zeta$ is the inhomogeneous
coordinate over $\PP_1$
the Riemann sphere, and $(\zeta,\eta)$ are the standard local coordinates on
$T$\PP$_1$ defined by $(\zeta,\eta)\rightarrow\eta\frac{d}{d\zeta}$. It 
must be real, with respect to the standard real structure on
$T$\PP$_1$
\be
\tau:(\zeta,\eta)
\mapsto(-\frac{1}{\bar{\zeta}},-\frac{\bar{\eta}}{\bar{\zeta}^2})\label{reality}\ee
and satisfy some non-singularity conditions \cite{Hb}.

The spectral curve of a 1-monopole positioned at $(x_1,x_2,x_3)$ is
called a star. It is
\be \eta-(x_1+ix_2)+2x_3\zeta+(x_1-ix_2)\zeta^2=0.\ee
\\

\noindent {\bf B}. Nahm data. 

Nahm data are anti-hermitian $n\times n$ matrices 
$(T_1,T_2,T_3)$ depending
on a real parameter $s\in[0,2]$ and satisfying the Nahm equation
\be\frac{dT_i}{ds}=\half\epsilon_{ijk}[T_j,T_k].\label{Nahmeqn}\ee 
The $T_i(s)$ are regular for $s\in(0,2)$ and have simple
poles at $s=0$ and $s=2$. The matrix residues of $(T_1,T_2,T_3)$ at each
pole form the irreducible $n$-dimensional representation of $su(2)$.

Nahm's equations (\ref{Nahmeqn}) are equivalent to a Lax pair. Hence, there
is an associated algebraic curve, which is, in fact, the spectral curve.
Explicitly, the \spc may be read off from the Nahm data as the
equation
\be
\mbox{det}(\eta+(T_1+iT_2)-2iT_3\zeta+(T_1-iT_2)\zeta^2)=0.
\label{lax}
\ee
 \\

\noindent {\bf C}. Based rational maps. 

A based rational map of degree $n$ from \PP$_1$
into \PP$_1$ sending $z=\infty$ to $z=0$ can be
given by
\be
r(z)=\frac{p(z)}{q(z)}
\ee
where $q(z)$ is a monic polynomial in $z$ of degree $n$ and
$p(z)$ is a polynomial in $z$ of degree less than $n$, which
has no factor in common with $q(z)$. 
We let $R_n$ denote the space of such rational maps.

It was conjectured in \cite{HMM} and proved by Bielawski \cite{Bi}, that for a rational map $p(z)/q(z)$ with well-separated poles $\beta_1,\ldots,\beta_n$ the corresponding monopole is
approximately composed of well-separated 1-monopoles with phases
$p(\beta_i)/|p(\beta_i)|$ located at the points $(x_1,x_2,x_3)$,
 where $x_1+ix_2=\beta_i$ and $x_3=\frac{1}{2}\log{|p(\beta_i)|}$.  This approximation applies only
when the values of the numerator at the poles is small compared to the
distance between the poles. 

The rational map provides a convenient parameterization of the moduli
space and with it it is possible to define the moduli space of strongly
centred monopoles precisely. A monopole with rational map $p(z)/q(z)$
is strongly centred if the roots of $q(z)$ sum to zero and the product
of $p(z)$ evaluated
at each of the roots of $q(z)$ is equal to unity. Thus, if we label the
roots of $q(z)$ as $\beta_1,\ldots,\beta_n$, as above, a monopole is strongly centred if
\bea \sum_{i}\beta_i&=&0\label{sc1}\\
    \prod_ip(\beta_i)&=&1\label{sc2}.\eea
The spectral curve of a strongly centred monopole has $a_1(\zeta)=0$.
The Nahm data of a strongly centred monopole are traceless.

Further asymptotic information can be derived from the rational
map. In \cite{AH} pp. 25-26, it is argued that for monopoles strung out in
well-separated clusters along, or nearly along, the $x_3$-axis, the
first term in a large $z$ expansion of the rational map $r(z)$ will be
$e^{2x+i\chi}/z^L$ where $L$ is the charge of the topmost cluster and
$x$ is its elevation above the plane.

It is possible to understand the action of the group $O(3)$ on the
spectral curves. The group of improper and proper rotations $O(3)$ has
a natural action on the
Riemann sphere. This is just the natural action on the sphere in
$\R^3$ written in terms of the inhomogeneous coordinate. This lifts to
an action on all of $T\PP_1$ and so gives
an action on spectral curves.

It is also possible to understand the
action of the rotation group on the Nahm data. The strongly centred Nahm data are an $\R^3\otimes sl(n,\C)$ valued function of $s$, which
transform under the rotation group $SO(3)$ as 
\begin{eqnarray}\underline{3}\otimes sl(\underline{n})
 &\cong&\underline{3}\otimes 
(\underline{2n-1}\oplus\underline{2n-3}\oplus
 \ldots \oplus \underline{3})\nonumber\\
&\cong&( \underline{2n+1}\oplus \underline{2n-1}\oplus
\underline{2n-3})\oplus \ldots \oplus (
\underline{5}\oplus \underline{3}\oplus\underline{1}).
\label{decomp}
\end{eqnarray}
where $\underline{r}$ denotes the unique irreducible $r$-dimensional
representation of $su(2)$. Thus, for example, for 3-monopoles the Nahm
data transform as the representation 
\be
(\underline{7} \oplus \underline{5} \oplus \underline{3}) \oplus
(\underline{5} \oplus \underline{3} \oplus \underline{1})
\label{Nahmrep}.\ee

The construction of the rational
map of a monopole requires a decomposition of space as
$\R^3\cong\R\times\C$ and only those spatial rotations and reflections which respect
this decomposition act on the space of rational maps. This means that
a rational map can be rotated about the $x_3$-axis and can be
reflected in the $x_1x_2$-plane, but other $O(3)$ transformations
are unknown.

\section{Symmetric rational maps}
\news
We can easily calculate the action of $I$ on the space of rational
maps, $R_n$. From \cite{HMM} the action of the reflection\footnote{In
\cite{HMM} (and some of our previous papers) this reflection is referred
to as inversion, but in this paper we follow the more standard notation
and reserve the term inversion for the operation (\ref{definv}).}
in the $x_1x_2$-plane, 
\be\sigma:(x_1,x_2,x_3)\mapsto(x_1,x_2,-x_3),\ee
on $R_n$ is 
\be \sigma:\frac{p(z)}{q(z)}\mapsto\frac{\sigma(p)(z)}{q(z)}\ee
where $\sigma(p)(z)$ is the unique degree $n-1$ polynomial such that
\be \sigma(p)(z)p(z)\equiv1\qquad \mbox{mod}\,q(z)\label{refl1}.\ee 
If the roots, $\beta_i$, of $q(z)$ are distinct a useful alternative way of
obtaining $\sigma(p)(z)$ is to notice that it is the unique polynomial
of degree less than $n$ such that 
\be \sigma(p)(\beta_i)p(\beta_i)=1\label{refl2}\ee
for all $i$.

The action of a rotation, $Rot_{\theta}$, about the $x_3$-axis is also
known,
 \be Rot_{\theta}:\frac{p(z)}{q(z)}\mapsto\frac{p(\lambda z)}{\lambda^{-n}q(\lambda z)}\qquad \lambda=e^{-i\theta}.\label{rot}\ee
The factor of $\lambda^{-n}$ ensures that $q(z)$ remains monic under the
rotation and so guarantees that the rotated rational map is strongly centred.
We will be interested in $Rot_{\pi}$; 
 \be Rot_{\pi}:\frac{p(z)}{q(z)}\mapsto\frac{p(-z)}{-q(-z)}.\label{rotpi}\ee

Since $I=\sigma\circ Rot_{\pi}$ we can act with inversion on an
element of $R_n$. This allows us to calculate the form of an
inversion invariant element of $R_3$. First we calculate the form of
the numerator, $q(z)$, of such an element. The action of the reflection
does not affect $q(z)$. This means that it must be invariant under
$Rot_{\pi}$. Since $Rot_{\pi}$ acts on $q(z)$ by
changing the sign of $z$ and the overall sign 
$ q(z)=z(z^2-\beta^2) $ 
for some $\beta\in\C$. The candidate inversion symmetric rational map is
thus
\be r(z)=\frac{az^2+bz+c}{(z-\beta)z(z+\beta)}\ee
for complex $a$, $b$ and $c$. From (\ref{refl1}) and (\ref{rotpi}) this is
inversion invariant iff 
\be p(-z)p(z)\equiv 1\qquad\mbox{mod}\;\, z^3-\beta^2z\ee
or equivalently, by (\ref{refl2}), we require
\bea p(\beta)p(-\beta)&=&1,\label{I1}\\
     p(0)^2&=&1.\label{I2}\eea
The first strong-centring condition (\ref{sc1}) is automatically
satisfied by $q(z)$. The second strong-centring condition
(\ref{sc2}) is given by
\be p(\beta)p(0)p(-\beta)=1.\ee
When combined with (\ref{I1}) and (\ref{I2}) this gives
$p(0)=1$ and thus $c=1$. Explicitly substituting $p(z)$ into (\ref{I1})
gives the condition
\be  b^2-a^2\beta^2=2a.\label{constraint}\ee
This defines a surface in $\C^3$ of two complex dimensions
corresponding to inversion symmetric strongly centred 3-monopoles.
The argument above assumes that the roots are distinct, that is that $\beta\not=0$. By using
(\ref{refl1}) it can be demonstrated that (\ref{constraint}) applies in
the $\beta=0$ case as well.
We denote the space of inversion symmetric,
strongly centred rational maps by $R_I$. 
\be
R_I=\left\{\frac{az^2+bz+1}{z(z^2-\beta^2)}\,:\,b^2-a^2\beta^2=2a\right\}.
\ee

We can find some particularly symmetric 1-parameter families of maps in
$R_I$. We impose $C_2$ symmetry around the $x_3$-axis by requiring
invariance under $Rot_{\pi}:z\mapsto-z$. This
means that $b=0$ and, hence, either $a=0$ or $a=-2/\beta^2$. The
$C_2$-symmetric rational maps in $R_I$ are then 
\bea
r_1(z)&=&\frac{1}{z(z^2-\beta^2)}\\[4pt]
r_2(z)&=&\frac{-\frac{2}{\beta^2}z^2+1}{z(z^2-\beta^2)}
\label{D2ratmaps}\eea
where $\beta\in\C$ in $r_1$ and $\beta\in\C^{\times}$ in $r_2$. They correspond to monopoles symmetric under inversion
and rotation by $\pi$ about the $x_3$-axis. If we impose a further
symmetry, that of reflection in the $x_1x_3$-plane, $\beta^2$ is
forced to be real and thus $\beta$ either purely real or purely imaginary.

In Section 3, we discussed how to relate the rational maps of
well-separated monopoles to the positions of those monopoles.
In this spirit, we can examine the rational maps $r_1$ and $r_2$ for
extreme values of $\beta$. In the case of $r_1$, if $\beta$
is large and real, it corresponds to three monopoles along the
$x_1$-axis positioned at the origin and $\pm\beta$. If $\beta$ is zero,
the rational map corresponds to the 3-monopole axisymmetric about the
$x_3$-axis. If it is large and imaginary, it corresponds to monopoles
along the $x_2$-axis positioned at the origin and $\pm i\beta$. 

In
the case of $r_2$ there are two distinct geodesics, one with
real $\beta$ and the other with imaginary $\beta$.
 If $\beta$ is large and real, it corresponds
to three monopoles along the
$x_1$-axis positioned at the origin and $\pm \beta$. For $\beta$ 
small and real it corresponds to monopoles     
 along the
$x_3$-axis positioned at the origin and $\pm\log\sqrt{\frac{2}{\beta^2}}$.
 The imaginary
$\beta$ geodesic is similar except the $x_2$-axis replaces the
$x_1$-axis. 

Of course it is possible to interpret these geodesics as
corresponding to low energy scattering events. They are events in
which two monopoles approach a third monopole positioned half way
between them, instantaneously form a torus and then separate out and
move along a line at right angles to the one along which they
previously moved.    

\section{Symmetric spectral curves}
\news
On $T\PP_1$ reflection is given
by
\be
\sigma:(\zeta,\eta)\mapsto(\frac{1}{\bar{\zeta}},-\frac{\bar{\eta}}{\bar{\zeta}^2})\ee
and the rotation $Rot_\pi$ by
\be Rot_{\pi}:(\zeta,\eta)\mapsto(-\zeta,-\eta).\ee 
Thus, since $I=\sigma\circ Rot_{\pi}$,
\be
I:(\zeta,\eta)\mapsto(-\frac{1}{\bar{\zeta}},\frac{\bar{\eta}}{\bar{\zeta}^2}).\ee
Reality requires a spectral curve to be invariant under
$\tau$, (\ref{reality}). Since $I\circ\tau:(\zeta,\eta)\mapsto(\zeta,-\eta)$ a charge $n$
spectral curve is inversion symmetric if all its terms are of even
degree in
$\eta$ if $n$ is even and of odd degree if $n$ is odd.

The spectral curve of an inversion symmetric 3-monopole must be of the
general form
\be
\eta^3+(c_{1}+c_{2}\zeta+r\zeta^2-\bar{c}_{2}\zeta^3+\bar{c}_{1}\zeta^4)\eta=0
\ee
where $c_i\in\C$ and $r\in\R$. The values of the $c_i$ and
$r$ are constrained by the non-singularity conditions satisfied
by the spectral curves of monopoles. This spectral curve is almost
identical in form to the one presented by Hurtubise in
\cite{hurt}. It differs in the overall factor of $\eta$. We
follow Hurtubise in rotating the general spectral curve so that it is
in the form
\be \eta^3+(a_1+a_2\zeta^2+a_1\zeta^4)\eta=0\label{D2spectralcurve}\ee
where $a_i\in\R$.
By choosing this standard orientation, we can observe the extra symmetries
automatically satisfied by inversion symmetric 3-monopoles. For
$a_1\not=0$ the spectral curve has a $\Z_2\times\Z_2$ symmetry
corresponding to rotations of $\pi$ about the three cartesian
axes. This means the inversion symmetric 3-monopole has a $D_2$ symmetry. For
$a_1=0$ there is a 
S$^1\times\Z^2$ symmetry. For the particular value of $a_2$ determined
by the non-singularity conditions this is the spectral
curve of the axisymmetric monopole. 

The spectral curve (\ref{D2spectralcurve}) is symmetric under inversion,
rotation of $\pi$ about the $x_3$-axis and reflection in the
$x_1x_3$-plane. It has the same symmetries as the rational maps
$r_1(z)$
and $r_2(z)$. This means that the non-singularity constraints
satisfied by $a_1$ and $a_2$ restrict them to, at most, 1-parameter families. 

Even without considering the non-singularity conditions, we have learned
a lot about inversion symmetric monopoles by considering 
the spectral curve. The rational maps $r_1(z)$ and
$r_2(z)$ indicate that there are inversion symmetric 3-monopoles
which are $C_2$ and reflection symmetric. Because all inversion
symmetric $n=3$ spectral curves can be rotated to the form
(\ref{D2spectralcurve}), the spectral curve demonstrates that
 all inversion symmetric 3-monopoles are
$D_2$-symmetric about some triplet of orthogonal axes. It is also
true that all 2-monopoles are $D_2$-symmetric about some triplet of
orthogonal axes.

\section{Symmetric Nahm data}
\news
To find the 1-parameter family of values of $a_1$ and $a_2$ we rely on
the Nahm data formulation and construct $D_2$ invariant Nahm data. The
construction for Nahm data invariant under a finite rotational symmetry
group was
introduced in \cite{HMM} and is discussed in \cite{HSa}. 
Nahm
data invariant under the $D_2$ transformation is given by
$$
T_1(s)=\frac{f_1(s)}{2}\left[\begin{array}{ccc}
0&\sqrt{2}i&0\\
\sqrt{2}i&0&\sqrt{2}i\\
0&\sqrt{2}i&0
\end{array}
\right],
\;\; 
T_2(s)=\frac{f_2(s)}{2}\left[\begin{array}{ccc}
0&\sqrt{2}&0\\
-\sqrt{2}&0&\sqrt{2}\\
0&-\sqrt{2}&0
\end{array}
\right],$$
\be
T_3(s)=\frac{f_3(s)}{2}\left[\begin{array}{ccc}
-2i&0&0\\
0&0&0\\
0&0&2i
\end{array}
\right].\label{Nd}
\ee
In the notation of (\ref{Nahmrep}) these invariant Nahm data
correspond to the $SO(3)$ invariant $\underline{1}$ and the
$SO(2)$ and $D_4$ invariant vectors in $\underline{5}$. 

Nahm's equations for these data become
\be \frac{df_1(s)}{ds}=f_2(s)f_3(s)\label{reduceeq}\ee
and its cyclic permutations. The corresponding spectral curve is
\be
{ \eta}^{3} +  \left( 
(f_1^2-f_2^2) + (2f_1^2+2f_2^2 - 4f_3^2) \zeta^2+ 
(f_1^2-f_2^2) \zeta^4\right) \eta=0
\ee
and so we can identify the constants $a_1=f_1^2-f_2^2$ and
$a_2=2(f_1^2+f_2^2-2f_3^2)$.
Equation (\ref{reduceeq}) and its cyclic permutations are
the Euler top equations, which are well known to be solvable
in terms of elliptic functions. We take the solution in the form
given by Dancer when
examining symmetric Nahm data for $SU(3)$
monopoles \cite{Dancer}
\be f_1(s)=-\frac{D\mbox{dn}_k(Ds)}{\mbox{sn}_k(Ds)},\;\;
     f_2(s)=-\frac{D}{\mbox{sn}_k(Ds)},\;\;
     f_3(s)=-\frac{D\mbox{cn}_k(Ds)}{\mbox{sn}_k(Ds)}.\ee
where $D$ is a constant. The sn$_k(s)$, cn$_k(s)$ and dn$_k(s)$ are, of
course, Jacobi elliptic functions. The parameter $k$ is the modulus of the Jacobi elliptic functions and
$0 \le k< 1$.  Details of the Jacobi elliptic
functions can be found in, for example, Whittaker and Watson
\cite{WW}. 

To determine $D$ we require that the data satisfy 
the boundary condition given in section 3. We must
examine the elliptic functions near $s=0$ and $s=2$. Near $s=0$ we have
sn$_k(Ds)\sim Ds$ whereas cn$_k(0)=1$ and dn$_k(0)=1$. The
functions $f_1(s)$, $f_2(s)$ and $f_3(s)$ all have simple poles at
$s=0$ with residues $-1$. It is now easy to verify that the
residue matrices are an irreducible 3-dimensional representation of
$su(2)$. The functions have another simple pole at $Ds=2K$ where $K$ is
the complete elliptic integral of the first kind with modulus $k$.
Again it is simple to check that the irreducible
representation boundary conditions are
satisfied at this pole also.
The functions are analytic for
$0<s<2K/D$. If we set $D=K$, the data are valid Nahm data.

By
substituting into the expressions for $a_1$ and $a_2$ and using the
standard elliptic function identities
$\mbox{sn}_k^2(u)+\mbox{cn}_k^2(u)=1$ and
$k^2\mbox{sn}_k^2(u)+\mbox{dn}_k^2(u)=1$, we obtain
\be a_1=-K^2k^2,\qquad\qquad
     a_2=-2K^2(k^2-2).\ee
The spectral curve
(\ref{D2spectralcurve}) is now
\be
\eta^3-K^2\left(k^2+2(k^2-2)\zeta^2+k^2\zeta^4\right)\eta=0.\label{11}\ee
Using the standard formula \cite{WW}
\be K=\int_0^{\half\pi}(1-k^2\mbox{sin}^2\phi)^{-\half}d\phi\ee
gives that when $k=0$, $K=\pi/2$ and so the spectral curve is
\be
\eta^3+\pi^2\zeta^2\eta=0,\ee
which is the spectral curve of a 3-monopole symmetric about the
$x_3$-axis. As $k\rightarrow 1$, $K\rightarrow\infty$ and the spectral
curve is asymptotic to the product of stars
\be \eta(\eta-K(1-\zeta^2))(\eta+K(1-\zeta^2))=0.\ee
This describes three well-separated monopoles located at
positions $(\pm K,0,0)$ and $(0,0,0)$. We note that the Nahm data
correspond to monopoles moving along the $x_1$-axis. We have
explicitly plotted surfaces of constant energy density for these
monopoles and they are discussed in Section 9.  

Recently one of us has used an $n$-dimensional generalization of the $D_2$-symmetric
Nahm data discussed in this Section to produce 1-parameter families of
spectral curves with interesting properties. Details may be found in \cite{S}. 

The spectral curve (\ref{11}) is very similar to the 2-monopole spectral curve
derived by Hurtubise \cite{hurt}. The easiest way to derive this spectral curve
is to consider 2-monopole Nahm data. Since a
2-monopole is always $D_2$-symmetric about some triplet of orthogonal
axes we choose an orientation and construct invariant Nahm data as
above. They are given by
\be T_1(s)=\frac{f_1(s)}{2}\left(\begin{array}{cc} 0& i \\
i&0\end{array}\right),\; T_2(s)=\frac{f_2(s)}{2}\left(\begin{array}{cc} 0& 1 \\
-1&0\end{array}\right),\; T_3(s)=\frac{f_3(s)}{2}\left(\begin{array}{cc} -i& 0 \\
0&i\end{array}\right) \label{Nd2}\ee
where the $f_1,f_2,f_3$ are the same as those defined earlier.
By (\ref{decomp}), 2-monopole Nahm data transforms under $SO(3)$ as
$\underline{5}\oplus\underline{3}\oplus\underline{1}$. As with
the 3-monopole Nahm data the $D_2$-symmetric Nahm data correspond to
the $SO(3)$ invariant $\underline{1}$ and the
$SO(2)$ and $D_4$ invariant vectors in $\underline{5}$.   

The 2-monopole spectral curve (\ref{lax})
is then
 \be
\eta^2-\frac{K^2}{4}\left(k^2+2(k^2-2)\zeta^2+k^2\zeta^4\right)=0,\ee
which is the one obtained by Hurtubise \cite{hurt} using different
methods. 
If this technology, based on obtaining the spectral curve from Nahm
data, had been
available to Hurtubise his task in \cite{hurt} would have
been simpler.

\section{The metric}
\news
We have constructed a 4-dimensional submanifold of the 3-monopole
moduli space. It is the fixed set of the inversion action on the
entire 3-monopole moduli space. The fixed point set of a finite group
action on a Riemannian manifold is always totally geodesic and so our
submanifold is a totally geodesic submanifold of $M_3^0$. Furthermore
this inversion action commutes with the action of $SO(3)$ and so,
since the entire 3-monopole moduli space is $SO(3)$ invariant, our
restricted moduli space has an $SO(3)$ invariant metric. In this
section we will show that this metric is the Atiyah-Hitchin metric.

We will do this by considering the metric on the space of Nahm
data. It has been proven by Nakajima \cite{Nak} that the moduli space
of Nahm data has the same metric as the corresponding moduli space of
monopoles. The moduli space of Nahm data is difficult to define since
it involves factoring the space of Nahm data by a set of $SU(n)$
transformations. It also requires the introduction of a fourth Nahm
matrix, $T_0(s)$. In any Nahm data calculation this is set to
zero by one of the $SU(n)$ transformations.

The metric on Nahm data corresponding to a tangent $Y=(Y_0,Y_1,Y_2,Y_3)$ is
given by 
\be \|Y_0,Y_1,Y_2,Y_3\|^2=-\int_0^2\sum_{i=0...3}\mbox{tr}(Y_i^2)ds
.\ee
Thus, for example, to calculate the metric on the Nahm data for
monopoles moving along the $x_1$-axis, we would first choose
a separation parameter $r$ 
which may be identified with monopole distance when the monopoles are
well-separated. We know, from above, that a suitable choice would be $r=kK$. 
We define the tangent vectors
\be Y_i=\frac{dT_i}{dr}\label{dtv} \ee
and the metric coefficient is then given by
\be g(r)=-\int_0^2\sum_{i=1..3}\mbox{tr}(Y_i^2)\ ds.\ee
Note that in general the tangent vectors can only be
defined by direct differentiation, as in equation (\ref{dtv}),
if certain identities hold between the matrices defining
the Nahm data, which ensure that such tangent vectors
are orthogonal to the gauge orbits (see \cite{Sb} for
an example). However, since in the present case each
Nahm matrix only involves a single function, the 
orthogonality to gauge orbits is guaranteed (see appendix).
By substituting from
(\ref{Nd}) we find that
\be
\sum_{i=1..3}\mbox{tr}(Y_i^2)=-2\left[\left(\frac{df_1}{dr}\right)^2
+\left(\frac{df_2}{dr}\right)^2+\left(\frac{df_3}{dr}\right)^2\right],
\label{1}\ee
which allows the metric to be calculated.

Using the rational map formulation a geodesic submanifold of $M_2^0$
space has been constructed, namely, the space of
inversion symmetric 3-monopoles.  By constructing a geodesic
submanifold of the monopole moduli space a
geodesic submanifold of the Nahm moduli space has been constructed. The Nahm data on this submanifold is given along
a radial geodesic by (\ref{Nd}) and, by $SO(3)$ transformation, this
geodesic generates the whole space of inversion symmetric, strongly
centred 3-monopole
Nahm data. We know how the Nahm data
transforms under $SO(3)$; it transforms like a vector in
$\underline{5}$. Furthermore, we know the Nahm data for a suitably oriented
2-monopole; it has been given above in (\ref{Nd2}). The 2-monopole
Nahm data transform the
same way under $SO(3)$ as the inversion symmetric 3-monopole
Nahm data. This means that the space of 2-monopole Nahm
data is identical to the space of 3-monopole Nahm data except the
matrices appearing in the 2-monopole case are the basis of
$\underline{2}$ and in the 3-monopole case they are the basis of
$\underline{3}$. Thus the two spaces are
identical apart from an overall factor.

This factor is easy to calculate. It is given by the ratio between the
traces of the squares of the matrices in the two cases. This factor is
four. The metric for 3-monopoles is four times that for
2-monopoles. Thus, for
example, to calculate the 2-monopole metric along the same geodesic as
in the 3-monopole case, tangents $Y_i$ can be calculated as before and in
this case
\be
\sum_{i=1..3}\mbox{tr}(Y_i^2)=
-\frac{1}{2}\left[\left(\frac{df_1}{dr}\right)^2
+\left(\frac{df_2}{dr}\right)^2+\left(\frac{df_3}{dr}\right)^2\right]\label{2}
.\ee
Since the 2-monopole metric is Atiyah-Hitchin, so is the metric on the
space of inversion symmetric 3-monopoles, except that it is rescaled
by a factor of four.

\section{The asymptotic metric}
\news
Recently, Gibbons and Manton \cite{GM2} have constructed an approximate  metric
for $n$ well-separated monopoles. It is instructive to construct this
metric for 3-monopoles with inversion symmetry. According to Gibbons
and  Manton,
 $n$-monopoles located at $\{\rhobf_{i}\}$ 
with phases $\{\theta_i\}$ have a metric
\be ds^2=g_{ij}d\rhobf_i\cdot d\rhobf_j+g_{ij}^{-1}(d\theta_i+{\bf
  W}_{ik}\cdot d\rhobf_k)(d\theta_j+{\bf
  W}_{jl}\cdot d\rhobf_l)\ee
where $\cdot$ denotes the usual scalar product on $\R^3$ vectors,
repeated indices are summed over and
\bea g_{jj}&=&1-\sum_{i\not=j}\frac{1}{\rho_{ij}}\qquad\mbox{(no sum over
  }j\mbox{)}\\
     g_{ij}&=&\frac{1}{\rho_{ij}}\,\;\qquad\mbox{(}i\not=j\mbox{)}\nonumber\\
     {\bf W}_{jj}&=&-\sum_{i\not=j}{\bf w}_{ij}\,\,\;\qquad\mbox{(no sum over
  }j\mbox{)}\nonumber\\
     {\bf W}_{ij}&=&{\bf w}_{ij}\,\,\qquad\mbox{(}i\not=j\mbox{)},\nonumber\eea
$\rhobf_{ij}=\rhobf_i-\rhobf_j$ and $\rho_{ij}=|\rhobf_{ij}|$. The
approximation is valid for
$\rho_{ij}\gg1$. The ${\bf w}_{ij}$ are Dirac potentials and are defined
by 
\be \mbox{curl}\ {\bf w}_{ij}=\mbox{grad} \frac{1}{{ \rho}_{ij}}\ee 
where the curl and grad operators are taken with respect to the $i$th
position coordinate $\rhobf_i$.

In the case of three monopoles that are
symmetric under inversion symmetry $\rhobf_1=\rhobf$, $\rhobf_2=0$ and  $\rhobf_3=-\rhobf$, and we write $\rho=|\rhobf|$.
Furthermore we require $d\theta_1=d\theta$, $d\theta_2=0$ and
$d\theta_3=-d\theta$. Denoting ${\bf w}_{12}={\bf w}_{23}$ by ${\bf
  w}$ so that ${\bf w}_{13}=\frac{1}{2}{\bf w}$, we have
\bea
g_{ij}&=&\frac{1}{\rho}\left(\begin{array}{ccc}\rho-\frac{3}{2}&1&\frac{1}{2}\\[4pt]
1&\rho-2&1\\[4pt] \frac{1}{2}&1&\rho-\frac{3}{2}\end{array}\right)\\[6pt]  
{\bf W}_{ij}&=&\frac{1}{2}\left(\begin{array}{ccc}-3{\bf w}&2{\bf
  w}&{\bf w}\\[4pt]2{\bf w}&-4{\bf w}&2{\bf w}\\[4pt]{\bf w}&2{\bf w}&-3{\bf
  w}\end{array}\right)\nonumber\eea
so
\be ds^2=4\left[\frac{1}{2}\left(1-\frac{2}{\rho}\right)d\rhobf\cdot
d\rhobf+\frac{1}{2}\left(1-\frac{2}{\rho}\right)^{-1}(-d\theta+2{\bf w}\cdot d\rhobf)^2\right].\ee
Up to the overall factor of four this is the asymptotic metric for two strongly centred
monopoles separated by a distance $\rho$. Note that in the 2-monopole
case $\rho$ is the separation of the two monopoles; whereas in the
3-monopole case it is the distance from the monopole at the origin to either
of the other two monopoles.

\section{Geodesic scattering}
\news
Since the metric on the moduli space of 
inversion-symmetric 3-monopoles is now known, we
can understand their low energy dynamics
by studying geodesics.
Their moduli space is
the Atiyah-Hitchin manifold 
which allows the known results about 2-monopole dynamics
\cite{AH,GM} to be
translated into results about the dynamics of 3-monopoles.

We have already discussed the right angle scattering geodesics
in terms of their rational maps.
From the point of view of the Atiyah-Hitchin submanifold
such a scattering process is associated with a geodesic
which passes over the 2-dimensional rounded cone submanifold \cite{AH}.
Since we have the Nahm data for these monopoles, we can construct the
monopole fields along such a geodesic, using the numerical ADHMN 
construction we introduced
previously \cite{HSa}. In Figure 1 we plot a surface of
constant energy density for various times along the geodesic
(corresponding to the elliptic modulus parameter values
$k=0.99,0.90,0.80,0.00$). In
Figure 1(a) we see three separated monopoles. As they approach, they
deform and merge to form a pretzel shape, Figure
1(b). It is interesting that the pretzel configuration closely
resembles the pretzel 3-skyrmion of Walet \cite{Walet}. Moving along
the geodesic, the monopole becomes more
ring-like, Figure 1(c). It instantaneously forms the torus, Figure
1(d), before separating out again, through the same configurations,
rotated through $\pi/2$, Figure 1(e-g).

There is a closed 2-monopole geodesic \cite{BM}, corresponding to two
orbiting monopoles, so we can immediately conclude that a
closed 3-monopole geodesic exists.
Following \cite{BM}, the value of the elliptic modulus $k$
for the rotating 3-monopole configuration is determined as
the root of the equation
\be
\int_0^{\half\pi}\frac{2k^2\sin^2\phi -
1}{\sqrt{1-k^2\sin^2\phi}}\ d\phi=0\ee
giving $k\approx 0.906$.

In Figure 2 we plot a surface of constant energy density for
this monopole. The monopole has been rotated so that the axis
of rotation (which is also shown) is in the plane of the page.
The monopole motion is a periodic orbit, rotating at constant
angular velocity about the shown axis, which is at an angle
of approximately $\pi/9$ to the vertical \cite{BM}.\\

\section{Conclusion}
\news
We have shown that the moduli space of inversion
symmetric 3-monopoles is an Atiyah-Hitchin submanifold
of the 3-monopole moduli space. Using this result
we have studied some geodesics in the 3-monopole
moduli space and examined the associated monopole
dynamics, including displaying energy density plots.

It is possible to apply inversion symmetry to $n$-monopoles for
$n>3$. However, the resulting geodesic submanifold has more than four
dimensions and so cannot be an Atiyah-Hitchin submanifold. We have
just learnt, however, that Bielawski \cite{Bi2} has succeeded in finding
geodesic Atiyah-Hitchin submanifolds of the $n$-monopole moduli space for each
$n$. These submanifolds correspond to $n$-monopoles with inversion
symmetry and the individual monopoles equally spaced along an axis.
Bielawski derives his result by considering the moment map
construction. 

One interesting property of these equally spaced monopoles
is clear from the asymptotic metric discussion in 
Section 8. Inversion symmetry requires that 
 $\rhobf_1=\rhobf$, $\rhobf_2=0$ and  $\rhobf_3=-\rhobf$. It is then
 neccessary to fix  $d\theta_1=d\theta$, $d\theta_2=0$ and
$d\theta_3=-d\theta$ in order to derive the asymptotic 2-monopole
metric. Similarly, for the asymptotic metric for $n$ equally spaced
monopoles to be the same, up to a factor, as that for 2-monopoles the
monopoles must be given $d\theta$'s proportional to their distance from
the origin. Thus, Bielawski's equally spaced monopoles have electric
charge proportional to their distance from the centre of mass.   
\\
\\
\\
\noindent{\bf Acknowledgements}

We would like to thank Nick Manton for many useful discussions
and also  Niels Walet, who inspired our interest 
in $D_2$ symmetric monopoles.
CJH thanks the EPSRC for a research studentship and the
British Council for a Chevening award.\\

\newpage
\appendix
\section{Appendix: Tangent Vectors}
\news
\renewcommand{\theequation}{A\arabic{equation}}
\ \indent 
In this brief appendix we show that the tangent vectors 
defined by direct differentiation of the Nahm data
are orthogonal to the gauge orbits, providing each Nahm matrix
involves only a single function.

Let $G$ be the group of analytic $SU(n)$-valued functions $g(s)$,
for $s\in[0,2]$, which are the identity at $s=0$ and $s=2$, and 
satisfy $g^t(2-s)=g^{-1}(s)$.
Then gauge transformations $g\in G$ act on $su(n)$-valued 
Nahm data as
\be
T_0\rightarrow gT_0g^{-1}-\frac{dg}{ds}g^{-1}, \hskip 20pt
T_i\rightarrow gT_ig^{-1}  \hskip 15pt i=1,2,3. 
\ee

We work in the gauge $T_0=0$, and define the tangent vectors
$Y_i$ by direct differentiation with respect to the
geodesic parameter, $r$ say, ie.
\be Y_i=\frac{dT_i}{dr}. \ee
We want to show that these tangent vectors are orthogonal
to the tangent vectors $W_i$ of the gauge orbits ie.
\be
<Y_i,W_i>=-\int_0^2\sum_{i=0..3}\mbox{tr}(Y_iW_i)\ ds=0.
\label{ipz}
\ee
To compute $W_i$ we consider the infinitesimal gauge
transformation given by
\be
g=1+\epsilon A
\ee
for $A\in su(n)$, and work to first order in $\epsilon$.
Since we have set $Y_0=0$ then we need only consider
$W_i$ for $i=1,2,3$. These tangent vectors are given by
\be
W_i=(gT_ig^{-1}-T_i)/\epsilon=[A,T_i].
\ee
Thus 
\be
<Y_i,W_i>=-\int_0^2\sum_{i=1..3}
\mbox{tr}(\frac{dT_i}{dr}[A,T_i])\ ds.
\ee
Now we make use of the fact that each of the three Nahm matrices
depends on a single function ie
\be
T_i=f_i M_i, \hskip 15pt i=1,2,3
\ee
where the $M_i$ are constant matrices.
Thus
\be
<Y_i,W_i>=-\int_0^2\sum_{i=1..3}\frac{1}{2}\frac{df_i^2}{dr}
\mbox{tr}(M_iAM_i-M_iM_iA)\ ds=0
\ee
by the cyclic property of the trace.
Thus the required result is proved.

\newpage
\noindent {\bf Figure Captions}

Figure 1(a-g). A surface of constant energy density at increasing
times. The corresponding values of the elliptic modulus are
(a) $k=0.99$, (b) $k=0.90$, (c) $k=0.80$, (d) $k=0.00$,
(e) $k=0.80$, (f) $k=0.90$, (g) $k=0.99$.

Figure 2. A surface of constant energy density for the
rotating 3-monopole, together with the axis of rotation.\\

\vspace*{2in}
two figures attached as separate gif files.



\newpage

\begin{thebibliography}{99}
\bibitem{AH} M.F. Atiyah and N.J. Hitchin,
\lq{\sl The geometry and dynamics of magnetic monopoles}\rq,
Princeton University Press, 1988.
\bibitem {BM} L. Bates and R. Montgomery, \lq{\sl Closed geodesics on
    the space of stable monopoles}\rq, Commun. Math. Phys. 118, 635 (1988). 
\bibitem{Bi} R. Bielawski, \lq{\sl Monopoles, particles and rational
    functions}\rq, Ann. Glob. Anal. Geom. 14 (to appear).
\bibitem{Bi2} R. Bielawski, \lq{\sl The existence of closed geodesics
    on the moduli space of $k$-monopoles}\rq, McMaster preprint, 1996.
\bibitem{Dancer} A.S. Dancer \lq{\sl Nahm's equations and
    hyperk\"ahler geometry}\rq, Commun. Math. Phys. 158, 545-568 (1993).
\bibitem{D} S.K. Donaldson, \lq{\sl Nahm's equations and the classification of
  monopoles}\rq, Commun. Math. Phys. 96, 387 (1984).
\bibitem{GM} G.W. Gibbons and N.S. Manton, \lq{\sl Classical and
    quantum dynamics of BPS monopoles}\rq,  Nucl. Phys. B274, 183
  (1986).
\bibitem{GM2} G.W. Gibbons and N.S. Manton, \lq{\sl The moduli space
    metric for well-separated BPS monopoles}\rq,  Phys. Lett. B356, 32 (1995).
\bibitem{HMM} N.J. Hitchin, N.S. Manton and M.K. Murray,
\lq{\sl Symmetric monopoles}\rq, Nonlinearity, 8, 661 (1995).
\bibitem{Hb} N.J. Hitchin, \lq{\sl On the construction of
    monopoles}\rq, Commun. Math. Phys. 89, 145 (1983).
\bibitem{HSa} C.J. Houghton and P.M. Sutcliffe, \lq{\sl Tetrahedral and
    cubic monopoles}\rq, Commun. Math. Phys. 180, 343 (1996).
\bibitem{hurt} J. Hurtubise, \lq{\sl SU(2) monopoles of charge 2}\rq,
  Commun. Math. Phys. 92, 195 (1983).
\bibitem{JT} A. Jaffe and C. Taubes, \lq{\sl Vortices and
monopoles}\rq, Boston, Birkh\"auser, 1980.
\bibitem{M2} N.S. Manton, \lq{\sl A remark on the scattering of BPS
    monopoles}\rq, Phys. Lett. 110B, 54 (1982).
\bibitem{N} W. Nahm, \lq{\sl The construction of all self-dual
multimonopoles by the ADHM method}\rq, in Monopoles in quantum field
theory, eds. N.S. Craigie, P. Goddard and W. Nahm, World Scientific,
1982.
\bibitem{Nak} H. Nakajima, \lq{\sl  Monopoles and Nahm's equations
    }\rq, in Sanda 1990, Proceedings, Einstein metrics and
    Yang-Mills connections. 
\bibitem{St} D. Stuart, \lq{\sl The geodesic approximation
 for the Yang-Mills-Higgs equation}\rq, Commun. Math. Phys. 166, 149 (1994).
\bibitem{S} P.M. Sutcliffe, \lq{\sl Symmetric monopoles and finite-gap
    Lam\'e potentials}\rq, J. Phys. A 29, 5187 (1996).
\bibitem{Sb} P.M. Sutcliffe, \lq{\sl The moduli space
metric for tetrahedrally symmetric
 4-monopoles}\rq, Phys. Lett. 357B, 335 (1995).
\bibitem{Walet} N.R. Walet, \lq{\sl Quantizing the B=2 and B=3
    skyrmion systems}\rq, preprint, UMIST-TP-96/1 (1996).
\bibitem{WW} E.T. Whittaker and G.N. Watson, \lq{\sl A course of modern
    analysis}\rq, Cambridge University Press, 1902.
\end{thebibliography}
\end{document}